\documentclass[aps,twocolumn,eqsecnum,unsortedaddress,showpacs]{revtex4}
\usepackage{amsmath,graphicx}

\def\sn{\mathrm{sn\,}}
\def\cn{\mathrm{cn\,}}
\def\arctanh{\mathrm{arctanh\,}}
\def\dn{\mathrm{dn\,}}

\begin{document}
\arraycolsep=2pt
\bibliographystyle{apsrev}
\title{Soliton generation of two-component Bose-Einstein condensates in
optical lattices}

\author{N.\ A.\ Kostov$^1$, V. Z. Enol'skii$^2$, V. S. Gerdjikov$^{2}$\footnote{Permanent address:
Institute for Nuclear Research and Nuclear Energy, Bulgarian
Academy of Sciences, Blvd. Tsarigradsko chaussee 72, 1784 Sofia,
Bulgaria.}, V. V. Konotop$^3$, M. Salerno$^2$}

\email[N. A. Kostov:]{nakostov@ie.bas.bg}

\email[V. Z. Enol'skii:]{vze@ma.hw.ac.uk} {} \email[V. S.
Gerdjikov:]{gerjikov@inrne.bas.bg}

 \email[V. V. Konotop:]{konotop@cii.fc.ul.pt}

 \email[M. Salerno:]{salerno@sa.infn.it}

\affiliation{$^1$Institute for Electronics, Bulgarian Academy of
Sciences, \\ Blvd. Tsarigradsko chaussee 72, 1784 Sofia, Bulgaria}

\affiliation{$^2$Dipartimento di Fisica ``E.R.Caianiello"
and Istituto Nazionale di Fisica della Materia (INFM),\\
Universit\'a di Salerno, via S. Allende, I84081 Baronissi (SA),
Italy;}

\affiliation{$^3$Centro de F\'{\i}sica Te\'orica e Computacional,
Universidade de Lisboa, \\ Av. Prof. Gama Pinto 2, Lisboa
1049-003, Portugal;}

\date{\today}

\begin{abstract}
Coupled nonlinear Schr\"{o}dinger equations (CNLS) with an
external elliptic function potential model a quasi
one--dimensional interacting two-component Bose-Einstein
condensate trapped in a standing light wave. New families of
stationary solutions of the CNLS with a  periodic potential are
presented and their stability studied by direct numerical
simulations. Some of these solutions allow reduction to Manakov
system. From a physical point of view these solutions can be
interpreted as exact Bloch states at the edge of the Brillouin
zone. Some of them are stable while others are found to be
unstable against modulations of long wavelength. The solutions
which are modulationally unstable are shown to lead to the
formation of localized ground states of the coupled BEC system.
\end{abstract}
 \maketitle

\section{Introduction}

Recent experiments on dilute-gas Bose-Einstein condensates
(BEC's) have generated great interest both from theoretical and
experimental points of view \cite{dgps99}. At ultra-low
temperatures the mean-field description for the macroscopic BEC
wave-function is constructed using Hartree-Fock approximation and
results in the Gross-Pitaevskii (GP) equation~\cite{dgps99}. The
latter one reduces to the one-dimensional (1D) nonlinear
Schr\"{o}dinger (NLS) equation with an external potential, in
particular, when the transverse dimensions of the condensate are
on the order of its healing length and its longitudinal dimension
is much longer than its transverse ones (see e.g.
\cite{clr00,konsa}). This is termed the {\it quasi-one
dimensional} (quasi-1D) regime of the GP equation.  In this
regime BECs remain phase--coherent, and the governing equations
are one-dimensional.

Several families of stationary solutions for the cubic NLS with
an elliptic function potential  have been recently presented in
Refs \cite{bgr00,bcdk01} and their stability has been examined
using analytic and numerical methods
\cite{bcdk01,ckr01,bcdkp01,CarSeg,EilEnoKo}.  In the quasi-1D
regime, the GP equations for two interacting BEC's reduce to
coupled nonlinear Schr\"{o}dinger (CNLS) equations with an
external potential \cite{mdfmm00,ba01} (see also Sec. 2 below).
When the scattering lengths of the two components, which
characterize inter-particle interactions are close to each other,
the CNLS equations reduce to the Manakov system with an external
potential.

In the present paper we study the stationary  two-component
solutions of the CNLS with an external potential. Several cases of
explicit solutions in terms of elliptic functions are analyzed and
their stability properties are studied numerically.  In
particular, we derive a set of stationary solutions with trivial
and non trivial  phases. We remark that some of the solutions
presented in this paper were also analyzed independently in Ref.
\cite{Deco}. In this work, however, all components of the CNLS
were assumed to be proportional to the same elliptic function. We
extend their results in the sense that we derive solutions of
CNLS whose components are expressed through different elliptic
functions. We also investigate the role played by these solutions
as possible initial states from which localized matter waves
(solitons) can be generated.

The paper is organized as follows. In section II we show how to
derive one dimensional equations for coupled BEC starting from the
original three dimensional problem using a multiple scale
expansion in the small amplitude limit. In section III we present
exact solution of the CNLS system with non trivial phases, while
in section IV we analyze their limits (trigonometric and
hyperbolic). In Section V we derive stationary solutions with
trivial phases for both proportional and non-proportional
components. In Section VI we discuss the physical properties of
the trivial phase solutions and show, by means of direct numerical
simulations, how these solutions may lead to the formation of
localized matter waves through the mechanism of  modulational
instability. In section VII we will discuss the present results
in comparison with previous work and  in Section VIII  the
conclusions of the paper  are briefly summarized.

\section{Basic equations}

 At very low temperatures, when the mean field approximation is
applicable, the evolution of two interacting BECs
can be described by two coupled GP equations $(j=1,2)$ (see
e.g. \cite{mdfmm00,ba01})
\begin{equation}
    \label{gp1}
    i\hbar \frac{\partial \Psi_j}{\partial t}=
    \left[-\frac{\hbar^2}{2m}\nabla^2+V_j({\bf r})+\frac{4\pi\hbar^2}{m}
    \sum_{l=1,2}a_{jl}|\Psi_l|^2\right]\Psi_j
\end{equation}
where atomic masses of the both components are assumed to be
equal, $V_j({\bf r})$ is an external trap potential, and $a_{ij}$
are the scattering lengths of the respective atomic interactions
(other notations are standard). In the case when it consists of
superposition of a magnetic trap providing cigar shape of the
condensate (elongated, say, along the $x$-axis) and an optical
trap inducing a lattice potential (which is assumed to be periodic
along the $x$-axes) one has ($j=1,2$)
\begin{eqnarray}
V_j({\bf r})&=&\frac{m}{2}\omega_{j}^2[\lambda^2x^2+y^2+z^2]+
U(\kappa x), \\
U(\kappa x)&=&U(\kappa (x+L)) \label{pot}
\end{eqnarray}

Although in the last expression we have imposed equality of the optical
potential for the both components, in  a general case one has to
distinguish the linear oscillator frequencies, $\omega_1$ and $\omega_2$,
when considering the two components corresponding to the different
magnetic moments. For example, in the experimental settings of
\cite{exper1} with $^{87}$Rb atoms
$\Omega=\frac{\omega_2}{\omega_1}=\sqrt{2}$. This fact has natural
implication on the resulting form of the effective system of coupled 1D
NLS equations. Indeed, different oscillator frequencies means that two
components are located in two different parabolic potentials, and thus
their effective densities are different when the number of atoms is equal.
As a consequence, even at approximately equal $s$-wave scattering lengths,
and thus for $a_{11}\approx a_{22}$, the two components will experience different
nonlinearities (the latter being proportional to the atomic densities).

Another important issue to be mentioned here is that a
cigar-shaped BEC can be viewed as a waveguide for matter waves. As
such it is characterized by its mode structure. As it is well
known (c.f. with the nonlinear optical waveguides~\cite{Ag}) the
intrinsic nonlinearity of a BEC results in the mode interaction
(and thus energy distribution among modes). If however the
nonlinearity is weak enough, the main state of the condensate can
be considered as a weakly modulated ground state of the underline
linear system. As it is clear that for a two-component BEC the
respective small parameter is the ratio between the density
energy of two-body interactions, $\frac{4\pi\hbar^2N_{j}
a_{jj}}{m a_{j}^3\lambda}$ (hereafter $a_j=\sqrt{\frac{\hbar}
{m\omega_j}}$ is the linear oscillator length and
$N_j=\int|\Psi_j|^2d{\bf r}$ is the number of atoms of the $j$th
component) to the density of the recoil energy
$\frac{\hbar^2}{2ma_j^2\lambda}$. In other words the small
parameter of the problem can be identified as $\epsilon=\frac{4\pi
Na_{11}}{a_1\lambda}\ll 1$ (where $N=N_1+N_2$ is the total number
of atoms). In this situation a self-consistent reduction of the
original 3D system (\ref{gp1}) to the effective 1D system of the
coupled equations can be provided by means of the multiple-scale
technique. Since the details of such a reduction have already
published elsewhere \cite{konsa,bks} for a single component BEC,
here we only outline the main steps.

Let us first introduce dimensionless variables
\begin{eqnarray*}
{\bf r}'=(x',{\bf r}_\bot^\prime)=\frac{{\bf r}}{a_0}, \qquad
t'=\frac 12 \omega_1 t, \qquad
\psi_j=\sqrt{\frac{2a_1^3}{N}}\Psi_j ,
\end{eqnarray*}
and rewrite Eqs. (\ref{gp1}) in the form
\begin{equation}\label{gp21}
 i\dot\psi_1  =\left[-\Delta'+
T({\bf r}')+ U(\kappa' x')  + g_{11}|\psi_1|^2 +
g_{21}|\psi_2|^2\right]\psi_1,
\end{equation}
\begin{equation}\label{gp22}
i \dot\psi_2  =\left[-\Delta'+\Omega^2 T({\bf r}')+ U(\kappa' x')
+ g_{12}|\psi_1|^2 + g_{22}|\psi_2|^2\right]\psi_2,
\end{equation}
where $\kappa'=\kappa a_1$, $\Omega=a_1^2/a_2^2$,
\[
T({\bf r}')=\lambda^2 x'{}^2 + r_\bot^{\prime 2}, \qquad U(\kappa'
x')=\frac{2}{\hbar \omega_1}V(\kappa' x'),
\]
and $g_{ij}=4\pi Na_{ij}/a_1$. Next consideration depends on the
magnitude of $\kappa$ (it is assumed that $U'(x)/U(x)=O(1)$. One
can distinguish three main cases:

(i) $\kappa\sim 1$. Then, the periodicity modifies the spectrum of the
  underline system introducing the effective group velocity dispersion.
  Resulting equations are just CNLS equations without periodic potential.
  This is the case similar to one considered in \cite{konsa} for the case of
  a single-component BEC.

 (ii) $\kappa' \ll \epsilon$ (say $\kappa \sim \epsilon^2$). In this case
  the periodic potential can be considered as smoothly varying, and
  somehow can be viewed as a limit of the case considered below.

(iii) $\kappa'=\alpha\epsilon$ where $\alpha\sim 1$. This is the case
  when the potential periodicity is of the order of the effective length
  of the nonlinearity. Below we concentrate on this last case.

To this end we consider two eigenvalue problems
\begin{eqnarray}
    \label{linear}
(-\Delta'+\lambda x'{}^2+r_\bot^{\prime 2} )\varphi_1 &= &
E_1\varphi_1 ,\nonumber \\
(-\Delta'+\Omega(\lambda x'{}^2+ r_\bot^{\prime 2}))\varphi_2 &= &
E_2\varphi_2
\end{eqnarray}
whose normalized ground states are well known:
\begin{eqnarray}
    \label{linsol1}
\varphi_1 &=& \frac{\lambda^{1/4}}{\pi^{3/4}}e^{-\frac 12(\lambda
  x'{}^2+r_\bot^{\prime 2})}, \nonumber \\
\varphi_2 &=& \frac{\Omega^{3/4}\lambda^{1/4}}{\pi^{3/4}}
  e^{-\frac{\Omega}{2}(\lambda x'{}^2+r_\bot^{\prime 2})}
\end{eqnarray}
and $E_j=\Omega^{j-1}(j+2)$.

The next steps are conventional for the multiple scale expansion (see
  e.g. \cite{konsa}). Namely we introduce scaled variables $x_n=\epsilon^n
  x'$, ${\bf r}_{n}=\epsilon^n{\bf r}_\bot^\prime$ and
  $t_n=\epsilon^nt'$ ($n=0,1,2,...$) which are considered as independent
  and look for the solution of (\ref{gp21}), (\ref{gp22}) in the form
\begin{eqnarray}
    \label{expan}
  \psi_j=\sqrt{\frac{1}{|g_{11}|\lambda^{1/2}}}\left(\epsilon
  \psi_j^{(1)}+\epsilon^2 \psi_j^{(2)}+\cdots\right)
\end{eqnarray}
with
\begin{eqnarray}
\psi_j^{(1)}=Q_j(x_1,t_2)\varphi_j (x_0,{\bf r}_0)e^{-iE_j
  t_0},\qquad j=1,2.
\end{eqnarray}
Here $Q_j(x_1,t_2)$ describes slow modulation of the background state
(\ref{linsol1}) due to the nonlinearity.

Substituting (\ref{expan}) in (\ref{gp21}), (\ref{gp22}), equating all
  terms at each of the $\epsilon$ orders, and excluding secular terms, in
  the order $\epsilon^3$ we obtain
\begin{equation}\label{gp31}
   i\frac{\partial Q_1}{\partial t_2} =-\frac{\partial^2 Q_1}{\partial
  x_1^2}+ V(\alpha x_1)Q_1+ \chi_{1}|Q_1|^2Q_1 + \chi|Q_2|^2Q_1,
\end{equation}
\begin{equation}\label{gp32}
i\frac{\partial Q_2}{\partial t_2} =-\frac{\partial^2
Q_2}{\partial
  x_1^2}+ V(\alpha x_1)Q_2+ \chi|Q_1|^2Q_2 + \chi_{2}|Q_2|^2Q_2,
\end{equation}
where
\begin{eqnarray*}
\chi_{1}&=& \mbox{sign} (g_{11})\int| \varphi_1|^4d{\bf
r}=\frac{\mbox{sign} (g_{11})}{2^{3/2}\pi^{3/2}}, \\
\chi&=& \frac{g_{12}}{|g_{11}|}\int|\varphi_1|^2|\varphi_2|^2d{\bf
r}=\frac{1}{\pi^{3/2}}\left(\frac{\Omega}
{\Omega+1}\right)^{3/2}\frac{a_{12}}{|a_{11}|}, \\
\chi_{2}&=& \frac{g_{22}}{|g_{11}|}\int|\varphi_2|^4d{\bf r}
=\frac{\Omega^{3/2}}{2^{3/2}\pi^{3/2}}\frac{a_{22}}{|a_{11}|},
\end{eqnarray*}
and its is taken into account that $a_{12}=a_{21}$. The system
(\ref{gp31}), (\ref{gp32}) is a subject of our main
interest.

\section{Stationary solutions with non-trivial phases}

After the change of notations: $t_2\to t $, $x_1\to x $, $\beta =-\chi $,
and $b_{1,2}=-\chi _{1,2} $ the system (\ref{gp31}), (\ref{gp32}) takes
the well known  form:
\begin{eqnarray}\label{eq:1}
i \frac{\partial Q_{1}}{\partial t}+ \frac{\partial^2
Q_{1}}{\partial x^2} &+& ( b_{1} |Q_{1}|^2 + \beta |Q_{2}|^2 )Q_1
\nonumber \\  && \qquad - V_0 \sn^2(\alpha x,k)Q_1=0, \\
\label{eq:1'} i i \frac{\partial Q_{2}}{\partial t}+
\frac{\partial^2 Q_{2}}{\partial x^2} &+& (\beta |Q_{1}|^2 + b_{2}
|Q_{2}|^2  )Q_2 \nonumber \\  && \qquad - V_0 \sn^2(\alpha
x,k)Q_2=0,
\end{eqnarray}

We restrict our attention to stationary solutions of these CNLS:
\begin{eqnarray}
Q_{j}(x,t)= q_{j}(x)\exp(-i\omega_{j}t+ i\Theta_{j}(x)+i\kappa
_{0,j} ),  \label{AManV}
\end{eqnarray}
where $j=1,2$, $\kappa _{0,j} $ are constant phases, $q_{j}$ and
$\Theta _j(x) $ are real-valued functions and
\begin{eqnarray} \label{Phase}
\Theta_{j}(x)={\mathcal{C}}_{j}\int_{0}^{x}
\frac{dx'}{q_{j}^{2}(x')}, \qquad j=1,2 ,
\end{eqnarray}
where ${\mathcal{C}}_{j}$, $j=1,2$ are constants of integration.

Following \cite{bcdk01} we refer to solutions in the cases
${\mathcal{C}}_{j}=0$ and ${\mathcal{C}}_{j}\neq 0$ as to trivial
and nontrivial phase solutions, respectively. We notice that
nontrivial phase solutions imply nonzero current of the matter --
it is proportional to $|q_j(x)|^2\Theta_{jx}={\mathcal{C}}_{j}$,
for each of the components -- along $x$-axis, and hence seem to
have no direct relation to present experimental setting for BECs
(remember that the condensate is confined to a parabolic trap).
Meantime a system of coupled NLS equations appears to be a general
model, having, for example, applications in nonlinear optics (see
e.g. \cite{Ag}). Bearing this in mind we consider both types of
solutions.

An appropriate class of periodic potentials to model the quasi-$1D$
confinement produced by a standing light wave is given by
\begin{eqnarray}\label{eq:pot}
V(\alpha x)=V_{0}\sn^2(\alpha x,k) ,
\end{eqnarray}
where $\sn(\alpha x,k)$ denotes the Jacobian elliptic sine
function with elliptic modulus $0\leq k\leq 1$. Then, substituting
the ansatz (\ref{AManV}) in Eqs. (\ref{gp31}), (\ref{gp32}) and
separating the real and imaginary part we get
\begin{equation}\label{sManV1}
q_{1}^{3} q_{1xx} + (b_1q_{1}^{2}+\beta q_{2}^{2}) q_{1}^{4} -
V_{0}\sn^{2}(\alpha x,k) q_{1}^{4} +\omega_{1} q_{1}^{4}=
{\mathcal{C}}_{1}^{2}  ,
\end{equation}
\begin{equation}\label{sManV2}
q_{2}^{3} q_{2xx} + (\beta q_{1}^{2}+b_2q_{2}^{2}) q_{2}^{4} -
V_{0}\sn^{2}(\alpha x,k) q_{2}^{4} +\omega_{2} q_{2}^{4} =
{\mathcal{C}}_{2}^{2}  .
\end{equation}
We seek solutions for $q_{j}^{2}$, $j=1,2$ as a quadratic
function of $\sn(\alpha x,k)$:
\begin{eqnarray}
q_{j}^{2}= A_{j}\sn^{2}(\alpha x,k)+B_{j} ,\qquad j=1,2 .
\label{AManV1}
\end{eqnarray}
Inserting (\ref{AManV1}) in (\ref{sManV1}), (\ref{sManV2}) and
equating the coefficients of equal powers of $\sn(\alpha x,k)$
results in the following relations among the solution parameters
$\omega_{j}$, $ \mathcal{{C}}_{j}$, $A_{j}$ and $B_{j}$ and the
characteristic of the optical lattice $V_{0}$, $\alpha$ and $k$:
\begin{equation}\label{A}
A_1=\frac{(b_2-\beta )W}{\Delta}, \qquad A_2=\frac{(b_1-\beta )W}
{\Delta} ,
\end{equation}
\begin{equation}\label{A'}
B_{j}=-\beta_{j} A_{j},\qquad {\mathcal{C}}_{j}^{2} = \alpha ^2
A_{j}^2 \beta_{j}(\beta_j-1)(1-\beta_{j} k^{2})
\end{equation}
\begin{eqnarray} \label{eq:12}
\omega_{j}&=&(1+k^{2})\alpha ^2  \\
\qquad &+& \frac{W}{\Delta}[\beta_1 b_1(b_2-\beta ) -\beta_2\beta
(\beta -b_1)] -k^2\alpha ^2\beta _j,\nonumber
\end{eqnarray}
where $j=1,2$ and
\begin{equation}
    \label{delta}
W=V_0-2\alpha^2k^2, \qquad
  \Delta=\chi_1\chi_2-\chi^2 =b_1b_2-\beta ^2.
\end{equation}

In order that our results (\ref{AManV1})--(\ref{eq:12}) are
consistent with the parametrization (\ref{AManV}), (\ref{Phase})
we must ensure that both $q_j(x) $ and $\Theta _j(x) $ are
real-valued; this means that $C_j^2\geq 0 $ and $q_j^2(x)\geq 0
$. An elementary analysis shows that this is true provided one of
the following pairs of conditions are satisfied:
\begin{eqnarray}\label{eq:A-b0}
&& \mbox{a)} \qquad A_j\geq 0, \quad \beta _j\leq 0, \quad j=1,2; \\
\label{eq:A_b0} && \mbox{b)} \qquad A_j\leq 0, \quad 1\leq \beta
_j\leq {1\over k^2}, \quad j=1,2;
\end{eqnarray}

Although our main interest is to analyze periodic solutions note
that the solutions $Q_{j}$ in (\ref{AManV}) {\em are not} always
periodic in $x$. Indeed, let us first calculate explicitly
$\Theta _j(x) $ by using the well known formula, see e.g.
\cite{as65}:
\begin{eqnarray*}
&& \int_{0}^{x}\frac{d u} {\wp(\alpha u)-\wp(\alpha v)}
\\
&& \qquad =\frac{1}{\wp'(\alpha v)}\left[2x\zeta(\alpha v)
+\frac{1}{\alpha } \ln \frac{\sigma(\alpha u-\alpha v)}
{\sigma(\alpha  u+\alpha v)} \right]
\end{eqnarray*}
where $\wp$, $\zeta$, $\sigma$ are standard Weierstrass functions.

In the case a) we replace $v $ by $iv_j $, set
$\sn^2(i\alpha v_j;k)=\beta_j<0$ and
\[ e_1=\frac13(2-k^2),\qquad  e_2=\frac13(2k^2-1),\qquad
e_3=-\frac13(1+k^2),
\]
and rewrite the l.h.s in terms of Jacobi elliptic functions:
\begin{eqnarray*}
&& \int_{0}^{x}\frac{du \;\sn^2(i\alpha v;k)
\sn^2(\alpha u;k) } {\sn^2(i\alpha v;k) -
\sn^2(\alpha u;k)}\\
&& \qquad = -\beta _jx -\beta _j^2 \int_{0}^{x} \frac{d
u\;}{\sn^2(\alpha u,k)-\beta _j}
\end{eqnarray*}
Skipping the details we find the explicit form of $\Theta _j(x) $:
\begin{eqnarray}\label{eq:Thet_j}
\Theta _j(x) &=& C_j \int_{0}^{x} \frac{d u}
{A_j({\sn}^2(\alpha u;k)-\beta _j)} \nonumber\\
&=& - \tau_j x +  \frac{i}{2} \ln \frac{\sigma(\alpha x+
i\alpha v_j)} {\sigma(\alpha x-i\alpha v_j)}, \\
\tau_j& = & i\alpha \zeta (i\alpha v_j)+ \frac{\alpha }{\beta_j}
\sqrt{-\beta_j(1-\beta_j)(1-k^2\beta_j)}. \nonumber
\end{eqnarray}

These formulae provide an explicit expression for the solutions
$Q_j(x,t) $ with nontrivial phases; note that for real values of
$v_j $ $\Theta _j(x) $ are also real.  Now we can find the
conditions under which $Q_j(x,t) $ are periodic.  Indeed, from
(\ref{eq:Thet_j}) we can calculate the quantities $T_j $
satisfying:
\begin{equation}\label{eq:T_j}
\Theta _j(x+T_j)-\Theta _j(x) = 2\pi p_j.
\end{equation}
Then $Q_j(x,t) $ will be periodic in $x $ with periods
$T_j=m_j\omega  $ if there exist pairs of integers $m_j $, $p_j$,
such that:
\begin{equation}\label{eq:period}
{m_j \over p_j} = -\pi \left[ \alpha v_j\zeta (\omega )+\omega
\tau_j \right]^{-1}, \qquad j=1,2.
\end{equation}
where $\omega $ (and  $\omega' $) are the half-periods of the
Weierstrass functions.

Of course the trivial phase solutions considered in the next sections are
always periodic functions of $x $.

We will list also solutions for two particular choices of $b_1 $,
$ b_2 $ and $\beta  $ which can be viewed as singular limits of
the generic case considered above. The first one is
\begin{equation}\label{eq:sing_1}
\beta ^2= b_1b_2, \qquad b_1 \neq b_2
\end{equation}
which corresponds to the case (\ref{eq:A-b0}).
Then the solution is given by:
\begin{eqnarray}\label{1-sin}
&& A_2 = -{ b_1 \over \beta  } A_1, \qquad V_0 = 2k^2\alpha ^2,
\nonumber\\
&&\omega_{1}= (\beta_1 -\beta_2) b_1 A_1 + (1+k^{2})\alpha ^2
-\alpha ^2 k^2 \beta _{1} ,  \\
&&\omega_{2}= (\beta_1 -\beta_2) \beta A_2 + (1+k^{2})\alpha ^2
-\alpha ^2 k^2 \beta _{2} ,  \nonumber\\
&&{\mathcal{C}}_{j}^{2} =  \alpha ^2 A_{j}^2 \beta_{j} (\beta_{j} -
1)(1-\beta_{j} k^{2}) ,\nonumber \\
\label{eq:12s} &&B_{j}=-\beta_{j} A_{j}, \qquad  j=1,2. \nonumber
\end{eqnarray}

The second particular case is the Manakov system; it corresponds
to the choice $b_1=b_2=b$. The result is
\begin{eqnarray}\label{}
&&\omega_{j}=b (\beta_1 A_{1}+\beta_2 A_2) + (1+k^{2})\alpha ^2
-\alpha ^2 k^2 \beta _{j} ,  \\
&&{\mathcal{C}}_{j}^{2} =  \alpha ^2 A_{j}^2 \beta_{j} (\beta_{j} -
1)(1-\beta_{j} k^{2}) ,\nonumber \\
\label{eq:12m} &&B_{j}=-\beta_{j} A_{j}, \qquad  V_{0}= b(
A_{1}+A_2) +2 k^2\alpha^2, \qquad j=1,2.  \nonumber
\end{eqnarray}

We remark that in the two-component CNLS  eqs. (\ref{sManV1}),
(\ref{sManV2}) the constants $b_1 $, $b_2 $ and $\beta  $ are
assumed to be positive. However in our considerations we do not
need this restrictions and our formulae are valid also for
negative values of $b_1 $, $b_2 $ and $\beta $.

\section{Limits of the non-trivial phase solutions}

\subsection{The limit $k\rightarrow 1$}

In this limit the elliptic functions reduce to hyperbolic
functions. Specifically, $\sn(x,k)=\tanh (x)$. Hence in this limit
and for repulsive BECs the solutions have the form
\begin{eqnarray}
&&q_{j}^{2}=A_{j}(\tanh^2(\alpha x)-\beta _{j}), \qquad j=1,2
,\nonumber \\
&&{\mathcal{C}}_{j}^{2} = -\alpha ^2 A_{j}^2 \beta_{j}(1-\beta_{j})^2,
\nonumber\\
&& \omega _j = \beta _1b_1A_1 + \beta _2\beta A_2 + (2-\beta _j)\alpha ^2.
\end{eqnarray}
The potential has only a single well or a single peak
$V(x)=-V_{0}\tanh^2(\alpha x)$. The consistency condition
(\ref{eq:A-b0}) in this case takes the form:
\begin{eqnarray}\label{eq:A-b1}
\mbox{a)} \qquad A_j\geq 0, \quad \beta _j\leq 0, \quad j=1,2;
\end{eqnarray}
while the second one (\ref{eq:A_b0}) degenerates and dissappears.

The same limit combined with the condition $\beta ^2=b_1b_2 $
leads  to:
\begin{eqnarray}\label{eq:ss}
&& q_j^2(x) = A_j (\tanh ^2(\alpha x) -\beta _j), \nonumber\\
&& \omega _1 = (\beta _1-\beta _2)b_1 A_1 + (2-\beta _1)\alpha ^2, \\
&& \omega _2 = (\beta _1-\beta _2)\beta  A_2 + (2-\beta _2)\alpha ^2,
\nonumber\\
&& C_j^2 = -\alpha ^2 A_j^2\beta _j(1-\beta _j)^2,\nonumber\\
&& A_2 = - {b_1  \over \beta  } A_1, \qquad V_0 = 2\alpha ^2,
\nonumber
\end{eqnarray}
and for the Manakov case $b_1=b_2=\beta =b $ we have:
\begin{eqnarray}\label{eq:man1}
&& q_j^2(x) = A_j (\tanh ^2(\alpha x) -\beta _j), \nonumber\\
&& \omega _j = b(\beta _1 A_1 +\beta _2 A_2) + (2-\beta _j)\alpha ^2,\\
&& C_j^2 = -\alpha ^2 A_j^2\beta _j(1-\beta _j)^2,\nonumber\\
&& V_0 = b(A_1+A_2)+2\alpha ^2, \nonumber
\end{eqnarray}
Then the nontrivial phases are equal to ($\beta _j<0 $):
\begin{equation}\label{eq:theta_j}
\Theta _j(x)= \alpha \sqrt{-\beta _j}x + \arctanh \left( {\tanh \alpha x
\over \sqrt{-\beta _j}}\right).
\end{equation}

\subsection{The trigonometric limit}
In the limit $k\rightarrow 0$, the elliptic functions reduce to
trigonometric functions and $V(x)=-V_{0}\sin^{2}(\alpha x)$. Then
\begin{eqnarray}
&&q_{j}^{2}=A_{j}(\sin^{2}(\alpha x)-\beta _{j}), \qquad j=1,2 ,\nonumber \\
&&\omega_{1}=\alpha^2 +\beta _1b_1A_1 +\beta _2\beta A_2, \nonumber \\
&&\omega_{2}=\alpha^2 +\beta _1\beta A_1 +\beta _2b_2 A_2, \nonumber \\
&&{\mathcal{C}}_{j}^{2} =\alpha^2 A_{j}^2 \beta_{j}(\beta_{j}-1),\\
&& A_1= {b_2-\beta \over b_1b_2-\beta ^2 }, \qquad A_2= {b_1-\beta
\over b_1b_2-\beta ^2 }, \nonumber
\end{eqnarray}
i.e., $V_0=b_1A_1+b_2A_2 $. The consistency conditions
(\ref{eq:A-b0}) and (\ref{eq:A_b0})  then take the form:
\begin{eqnarray}\label{eq:A-b2}
\mbox{a)} \qquad A_j>0, \quad \beta _j<0, \quad j=1,2; \\
\label{eq:A_b2} \mbox{b)} \qquad A_j<0, \quad \beta _j\geq 1,
\quad j=1,2;
\end{eqnarray}

If we assume $\beta ^2=b_1b_2 $ then
\begin{eqnarray}
&&q_{j}^{2}=A_{j}(\sin^{2}(\alpha x)-\beta _{j}), \qquad j=1,2 ,\nonumber \\
&&\omega_{j}=\alpha^2 + (-1)^{j+1} (\beta _1-\beta _2)b_1A_1 ,\nonumber \\
&&{\mathcal{C}}_{j}^{2} =\alpha^2 A_{j}^2 \beta_{j}(\beta_{j}-1),\\
&& A_2= -{b_1\over \beta  } A_1, \qquad V_0=0,    \nonumber
\end{eqnarray}
and in the Manakov case $b_1=b_2=\beta =b $ we have:
\begin{eqnarray}
&&q_{j}^{2}=A_{j}(\sin^{2}(\alpha x)-\beta _{j}), \qquad j=1,2 ,\nonumber \\
&&\omega_{j}=\alpha^2 + b(\beta _1A_1+\beta _2A_2),\nonumber \\
&&{\mathcal{C}}_{j}^{2} =\alpha^2 A_{j}^2 \beta_{j}(\beta_{j}-1),\\
&& V_0=b(A_1+A_2). \nonumber
\end{eqnarray}
Therefore the phase integral (\ref{Phase}) equals ($\beta_j<0 $):
\begin{eqnarray}
\Theta_{j}=-\arctan \left(\sqrt{ \frac{1-\beta _{j}}{-\beta _{j}}
}\tan(\alpha x)\right) .
\end{eqnarray}

\section{Trivial phase solutions.}

In this section we consider solutions of (\ref{eq:1}), (\ref{eq:1'}) with
trivial phase, i.e.  $C_1=C_2=0 $:
\begin{equation}\label{eq:2}
 Q_j(x,t) = e^{-i\omega_j t +i\kappa _{0,j}} q_j(x), \qquad j=1,2,
\end{equation}
and we will look for different possible choices for the functions $q_1(x)
$ and  $q_2(x) $. This type of solutions are more flexible and in certain
cases survive reductions of the constants $\beta ^2=b_1b_2 $ or the limit
to the Manakov case:  $b_1=b_2=\beta  $.  They are also relevant for
processes in BEC and nonlinear optics \cite{Ag}.

In the following we shall consider the $q_i(x)$ to be expressed
in terms of Jacobi elliptic functions, i.e. we assume
the following ansatz: $q_i(x)= \gamma_i J_i(x)$, with
$J_i(x), i=1,2 $ being one of the Jacobi elliptic function
$\sn(\alpha x,k) $, $\cn(\alpha x,k) $ or $\dn(\alpha x,k) $ and
$\gamma_i$ specifying both the real amplitudes and the constant
phases in (\ref{eq:2}).  Note that the CNLS (\ref{eq:1}),
(\ref{eq:1'}) possesses the gauge invariance $Q_j\rightarrow Q_j
e^{-i \kappa _{0,j}}$. This allows one to fix up conveniently the
initial phases of both $Q_j(x) $. In most of the following
examples we have made this choice by requiring that $\gamma _j^2>0
$.  Direct substitution of the above ansatz into Eqs.
(\ref{eq:1}), (\ref{eq:1'}) provides a set of algebraic equations
for the parameters whose solutions furnish exact ground states of
the coupled BEC system.

{\bf Case 1.} We start with
\begin{equation}\label{eq:2'}
q_1(x)=\gamma_1 \sn(\alpha x,k), \qquad q_2(x)=\gamma_2 \cn(\alpha
x,k),
\end{equation}

The functions in (\ref{eq:2}) are
solutions of (\ref{eq:1}) provided the constants satisfy the
relations:
\begin{eqnarray}\label{eq:3}
&& b_1 \gamma_1^2 -\beta\gamma_2^2 -W=0,\nonumber \\
&& \beta \gamma_1^2 -b_2 \gamma_2^2 - W=0, \\
&& \beta \gamma_2^2 + \omega_1 - \alpha^2 (k^2+1) =0,
\nonumber \\
&& b_2 \gamma_2^2 + \omega_2 - \alpha^2 =0. \nonumber
\end{eqnarray}
where for convenience we have introduced
\begin{equation}\label{eq:W}
W= V_0 -2k^2\alpha^2.
\end{equation}

From this system we can determine $4$ of the constants in terms of
the others. Let us split these  constants into two groups. The
first one:
\[ G_1 \simeq \{b_1, \quad b_2, \quad \beta, \quad W, \quad \alpha, \quad
k, \} \]
consists of constants determining the equations and the potential
and we assume they are fixed. The second group of constants
\[ G_2 \simeq \{ \omega_1, \quad  \omega_2, \quad \gamma_1, \quad \gamma_2,
\}
\]
characterize the corresponding soliton solution. Next we solve
(\ref{eq:3}) and express the constants $G_2$ in terms of $G_1$.
If $\beta^2\neq b_1b_2 $ we get the result:
\begin{eqnarray}\label{eq:4}
&& \omega_1 = -{ \beta (b_1-\beta) \over \beta^2 -b_1b_2}W +
\alpha^2 (k^2+1),  \nonumber \\
&& \omega_2 = -{ b_2 (b_1-\beta) \over (\beta^2 -b_1b_2)} W
+ \alpha^2 ,\\
&& \gamma_1^2 = { (\beta  -b_2)  \over \beta^2-b_1b_2 }W,  \qquad
\gamma_2^2 = { (b_1 -\beta) \over \beta^2 -b_1b_2 }W, \nonumber
\end{eqnarray}

The constraints on the constants $\gamma_1^2>0$ and $\gamma_2^2>0$
can be ensured in two ways:
\begin{equation}\label{eq:4a}
\left\{ \begin{array}{cc}
W<0 , \qquad & b_1>\beta >b_2,  \\
W>0 , \qquad & b_1<\beta <b_2, \end{array} \right.
\end{equation}

The case when $\beta ^2=b_1b_2 $ fixes up $W $ by
\begin{equation}\label{eq:V_0}
W=0,
\end{equation}
and then
\begin{eqnarray}\label{eq:4b}
&& \gamma _2^2 = \sqrt{{b_1\over b_2}} \gamma _1^2,  \nonumber\\
&&\omega _1 = \alpha ^2 (k^2+1) -b_1\gamma _1^2,\qquad \omega
_2=\alpha ^2-b_2\gamma _2^2.  \end{eqnarray}

{\bf Case 2.} Here
\begin{equation}\label{eq:2.2}
q_1(x)=\gamma_1 \sn(\alpha x,k), \qquad q_2(x)=\gamma_2 \dn(\alpha
x,k),
\end{equation}

The functions in (\ref{eq:2.2}) are  solutions of (\ref{eq:1}) provided
the constants satisfy the relations:
\begin{eqnarray}\label{eq:3.2}
&& b_1 \gamma_1^2 -k^2 \beta\gamma_2^2 -W=0,\nonumber \\
&& \beta \gamma_1^2 - k^2 b_2 \gamma_2^2 -W=0, \\ &&
\beta \gamma_2^2 + \omega_1 - \alpha^2 (k^2+1) =0, \nonumber \\ && b_2
\gamma_2^2 + \omega_2 - \alpha^2 k^2 =0. \nonumber
\end{eqnarray}
The solution of eq. (\ref{eq:3.2}) gives:
\begin{eqnarray}\label{eq:4.2}
&& \gamma _1^2 = {\beta -b_2  \over \beta ^2-b_1b_2 } W, \qquad
\gamma _2^2 = {b_1-\beta \over k^2(\beta ^2-b_1b_2 )} W,  \nonumber\\
&& \omega _1 = {\beta (\beta -b_1)  \over k^2(\beta ^2-b_1b_2) } W +
\alpha ^2 (k^2+1), \\
&& \omega _2 = {b_2(\beta -b_1)  \over k^2(\beta ^2-b_1b_2) } W +
\alpha ^2 k^2.\nonumber
\end{eqnarray}

The case $\beta ^2=b_1b_2 $ also fixes up $W $ by (\ref{eq:V_0})
and then
\begin{eqnarray}\label{eq:4c}
&& k^2 \gamma _2^2 = \sqrt{{b_1\over b_2}} \gamma _1^2,  \\
&&\omega _1 = \alpha ^2 (k^2+1) -{b_1 \over k^2}\gamma _1^2,\qquad
\omega _2=\alpha ^2- \sqrt{b_1b_2} \gamma _1^2.\nonumber
\end{eqnarray}

The solutions in cases 1 and 2 exclude the possibility to have
$b_1=b_2 $ and $\gamma _j^2>0 $.  The simplest way to see that is
to check that for $b_1=b_2 $ we have $\gamma _1^2+\gamma_2^2=0 $
from (\ref{eq:4}) and $\gamma _1^2+k^2\gamma_2^2=0 $ from
(\ref{eq:4.2}); i.e. either $\gamma _1^2 $ or $\gamma _2^2 $ must
be negative.

{\bf Case 3.} Here
\begin{equation}\label{eq:2.3}
q_1(x)=\gamma_1 \cn(\alpha x,k), \qquad q_2(x)=\gamma_2 \dn(\alpha
x,k),
\end{equation}

The functions in (\ref{eq:2.3}) are  solutions of (\ref{eq:1})
provided the constants satisfy the relations:
\begin{eqnarray}\label{eq:3.3}
&& b_1 \gamma_1^2 +k^2 \beta\gamma_2^2 + W=0,\nonumber \\
&& \beta \gamma_1^2 + k^2 b_2 \gamma_2^2 + W=0, \\ &&
b_1 \gamma_1^2 + \omega_1 +\beta \gamma _2^2 - \alpha^2  =0,
\nonumber \\
&& \beta  \gamma_1^2 + b_2\gamma _2^2 + \omega_2 - \alpha^2 k^2 =0.
\nonumber
\end{eqnarray}
The solution of eq. (\ref{eq:3.3}) gives:
\begin{equation}\label{eq:4.3}
\gamma _1^2 = { b_2-\beta   \over \beta ^2-b_1b_2 } W, \qquad
\gamma_2^2 = {b_1-\beta \over k^2(\beta ^2-b_1b_2 )} W,
\end{equation}
\begin{equation}\label{eq:4.3a}
 \omega _1 =\alpha ^2 + {\beta (\beta -b_1)  \over k^2(\beta^2-b_1b_2) }
W +{b_1(\beta -b_2)  \over \beta^2-b_1b_2 } W ,
\end{equation}
\begin{equation}\label{eq:4.3b}
\omega _2 = k^2 \alpha ^2 + {b_2 (\beta -b_1)\over k^2 (\beta^2
-b_1b_2) } W +{\beta (\beta -b_2)  \over \beta^2-b_1b_2 } W ,
\end{equation}

This solution {\em allows} the possibility to have $b_1=b_2=b $. It
reduces to
\begin{eqnarray}\label{eq:5.3}
&& \gamma _1^2 = -{W  \over \beta +b },\qquad
\gamma _2^2 = -{W \over k^2(\beta +b )}, \\
&&\omega _1 = \alpha ^2 -\gamma _1^2 \left( b + {\beta \over k^2
}\right), \qquad \omega _2 = \alpha ^2k^2 - \gamma _1^2 \left(
\beta  + {b\over k^2 }\right), \nonumber
\end{eqnarray}
Obviously to have $\gamma _j^2>0 $ we need to require that
\begin{equation}\label{eq:6.3}
W< 0.
\end{equation}

{\bf Case 4. } We put:
\begin{equation}\label{eq:c-7}
q_1(x)=\gamma _1 \dn (\alpha x,k) ,\qquad q_2(x)=\gamma _2 \sn
(\alpha x,k) .
\end{equation}
Then the corresponding sets of parameters satisfy:
\begin{equation}\label{eq:c-7.1}
\gamma _1^2 = {(\beta -b_2 )W \over k^2( \beta ^2-b_1b_2) },
\qquad \gamma _2^2 = {(\beta -b_1 )W \over \beta ^2-b_1b_2 },
\end{equation}
\begin{eqnarray}\label{eq:c-7.1'}
\omega _1 &=& \alpha ^2k^2  - { b_1(b_2-\beta )W \over k^2 (\beta
^2-b_1b_2) }, \\
\omega _2 &=& \alpha ^2(k^2 +1) - { \beta (b_2-\beta )W \over k^2
(\beta ^2-b_1b_2) }, \nonumber
\end{eqnarray}
The subcase $b_1=b_2=b $ is impossible since from Eq.
(\ref{eq:c-7.1}) there follows $k^2\gamma _1^2 + \gamma _2^2=0 $.
It is possible however to put $\beta ^2=b_1b_2 $ in which case:
\begin{eqnarray}\label{eq:c-7.2}
\gamma _2^2&=& k^2 \sqrt{{b_1  \over b_2 }} \gamma _1^2, \\
\omega _1 &=& \alpha ^2k^2 - b_1 \gamma _1^2, \\
\omega _2 &=&\alpha ^2 (k^2+1) - \sqrt{b_1b_2}\gamma _1^2.
\end{eqnarray}

{\bf Case 5. }Let now:
\begin{equation}\label{eq:c-8}
q_1(x)=\gamma _1 \dn (\alpha x,k) ,\qquad q_2(x)=\gamma _2 \cn
(\alpha x,k) .
\end{equation}
Then the corresponding sets of parameters satisfy:
\begin{equation}\label{eq:c-8.1}
\gamma _1^2 = -{(\beta -b_2 )W \over k^2(\beta ^2-b_1b_2) },
\qquad \gamma _2^2 = -{(\beta -b_1 )W \over \beta ^2-b_1b_2 },
\end{equation}
\begin{equation}\label{eq:c-8.1'}
\omega _1 =\alpha ^2 k^2 + {\beta (\beta -b_1)W \over \beta
^2-b_1b_2} + {b_1(\beta -b_2)W \over k^2(\beta ^2-b_1b_2)},
\end{equation}
\begin{equation}\label{eq:c-8.1''}
\omega _2 =\alpha ^2 + {b_2(\beta -b_1)W \over \beta ^2-b_1b_2} +
{\beta (\beta -b_2)W \over k^2(\beta ^2-b_1b_2)}.
\end{equation}

The subcase $b_1=b_2=b $ simplifies further (\ref{eq:c-8.1}) -
(\ref{eq:c-8.1''}) to
\begin{equation}\label{eq:c-8.2}
\gamma _1^2=-{W  \over k^2(\beta +b) }, \qquad \gamma _2^2=-{W
\over \beta +b },
\end{equation}
\begin{equation}\label{eq:c-8.2'}
\omega _1 =\alpha ^2 k^2 + \left( \beta + {b  \over k^2 } \right)
{W \over \beta +b} ,
\end{equation}
\begin{equation}\label{eq:c-8.2''}
\omega _1 =\alpha ^2 + \left( b+ {\beta \over k^2 } \right) {W
\over \beta +b} .
\end{equation}
Here again it is natural to consider $W<0 $.

Let us finally consider three more cases in which the two components
are proportional: $q_1(x)=\gamma q_2(x) $ and $q_1(x) $ is one of the
three functions $\sn (\alpha x,k) $, $\cn (\alpha x,k) $ or $\dn (\alpha
x,k) $. Such an ansatz imposes on the system (\ref{eq:1}), (\ref{eq:1'})
the compatibility condition

\begin{equation}\label{eq:gamma}
\gamma ^2 (\beta -b_2) +b_1-\beta +\omega _1-\omega _2=0
\end{equation}
If (\ref{eq:gamma}) is fullfilled the system (\ref{eq:1}),
(\ref{eq:1'}) reduces effectively to the one-component case, which has been
already studied; see also Section VII below.

{\bf Case 6. } We choose:
\begin{equation}\label{eq:c-4}
q_1(x)=\gamma _1 \sn (\alpha x,k) ,\qquad q_2(x)=\gamma _2 \sn
(\alpha x,k) .
\end{equation}
Then the corresponding sets of parameters satisfy:
\begin{eqnarray}\label{eq:c-4.1}
\gamma _1^2 = {(\beta -b_2 )W \over \beta ^2-b_1b_2 }, \qquad
\gamma _2^2 = {(\beta -b_1 )W \over \beta ^2-b_1b_2 }, \\
\omega _1 =\omega _2 =\alpha ^2(k^2+1). \nonumber
\end{eqnarray}
The subcase $b_1=b_2=b $ simplifies further (\ref{eq:c-4.1}) to
\begin{equation}\label{eq:c-4.2}
\gamma _1^2=\gamma _2^2={W  \over \beta +b }, \qquad \omega _1
=\omega _2 =\alpha ^2(k^2+1).
\end{equation}
Here it is natural to consider $W>0 $.

{\bf Case 7. } Assume:
\begin{equation}\label{eq:c-5}
q_1(x)=\gamma _1 \cn (\alpha x,k) ,\qquad q_2(x)=\gamma _2 \cn
(\alpha x,k) .
\end{equation}
Then the corresponding sets of parameters satisfy:
\begin{eqnarray}\label{eq:c-5.1}
\gamma _1^2 = -{(\beta -b_2 )W \over \beta ^2-b_1b_2 }, \qquad
\gamma _2^2 = -{(\beta -b_1 )W \over \beta ^2-b_1b_2 }, \\
\omega _1 =\omega _2 =\alpha ^2 +W. \nonumber
\end{eqnarray}
The subcase $b_1=b_2=b $ simplifies further (\ref{eq:c-5.1}) to
\begin{equation}\label{eq:c-5.2}
\gamma _1^2=\gamma _2^2=-{W  \over \beta +b }, \qquad \omega _1
=\omega _2 =\alpha ^2+W.
\end{equation}
Unlike case 6, here it is natural to consider $W<0 $.

{\bf Case 8. }Let here:
\begin{equation}\label{eq:c-6}
q_1(x)=\gamma _1 \dn (\alpha x,k) ,\qquad q_2(x)=\gamma _2 \dn
(\alpha x,k) .
\end{equation}
Then the corresponding sets of parameters satisfy:
\begin{eqnarray}\label{eq:c-6.1}
\gamma _1^2 = {(\beta -b_2 )W \over k^2(\beta ^2-b_1b_2) }, \qquad
\gamma _2^2 = {(\beta -b_1 )W \over k^2(\beta ^2-b_1b_2) }, \\
\omega _1 =\omega _2 =\alpha ^2k^2 + {W \over k^2}. \nonumber
\end{eqnarray}
The subcase $b_1=b_2=b $ simplifies further (\ref{eq:c-6.1}) to
\begin{equation}\label{eq:c-6.2}
\gamma _1^2=\gamma _2^2=-{W  \over k^2(\beta +b) }, \qquad \omega
_1 =\omega _2 =\alpha ^2k^2 +{W \over k^2}.
\end{equation}
Here again it is natural to consider $W<0 $.

\section{Modulational instability of the trivial phase solutions and
localized matter waves generation}\label{sec:4} To discuss the
stability of the above solutions we shall adopt a physical point.
To this end we remark that all the trivial phase solutions, are
periodic functions of period twice the period of the lattice
(recall that the period $a$ of potential in Eq(\ref{eq:pot}) is
$a=2K(k^2)/\alpha $, where  $K(k^2)$ is the complete elliptic
integral of the first kind). The corresponding wave-number of
these solutions is $\mathcal{K}=\pi/a$ which is  just the boundary
of the Brillouin zone of the uncoupled periodic linear system.
Moreover, one can easily check that each component $q_i(x)$,
$i=1,2,$ satisfy the Bloch condition
\begin{equation}
\label{bloch} q_i(x+R_n)= e^{(i\mathcal{K} R_n)} q_i(x), \qquad
R_n=n a,\qquad n\in N,
\end{equation}
i.e. the trivial phase solutions  are exact {\it nonlinear} Bloch
states (note that also in the nonlinear case  the concept of a
Bloch state is well defined by Eq. (\ref{bloch}). Although
nonlinearity does not compromise Bloch property (this being a
direct consequence of the translation invariance of the lattice),
it can drastically influence the stability of the states trough a
modulational instability mechanism.
\begin{figure}[h]
\includegraphics[scale=0.45,clip]{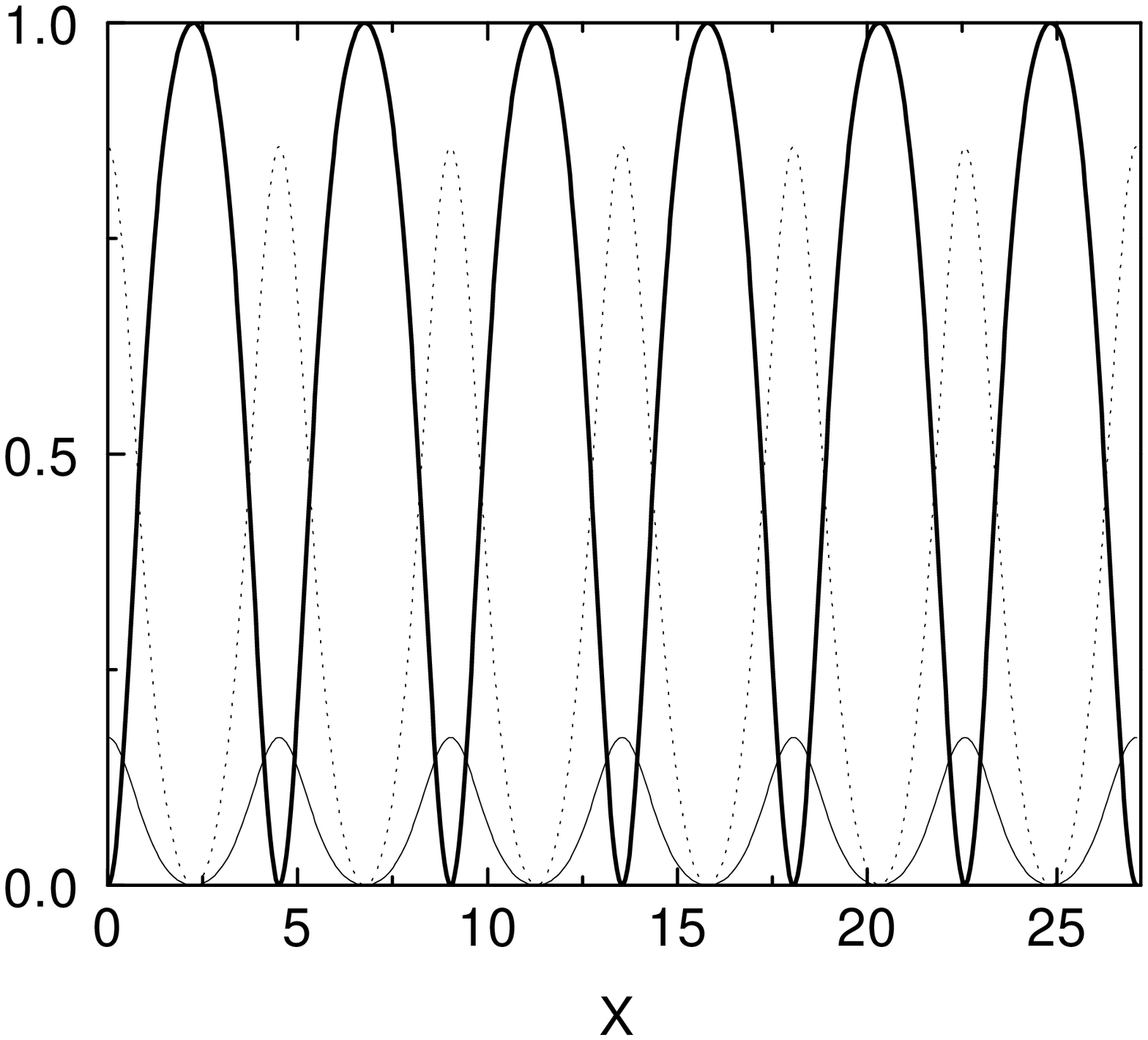}

\caption{Initial profile of stable cn-cn solution plotted against
the potential profile (thick curve). The continuous and dotted
thin curves denote, respectively, the modulo square of $q_1$ and
$q_2$. The parameters are fixed as: $k^2=0.8$, $V_0=1$, $\alpha=1$,
$\beta=.5$, $b_1=1.0$, $b_2=0.6$. The amplitudes of the two
components are $\gamma_1=-0.414039,
\gamma_2=0.92582$.\label{FIG1}}
\end{figure}

\begin{figure}[h]
\includegraphics[scale=0.45,clip]{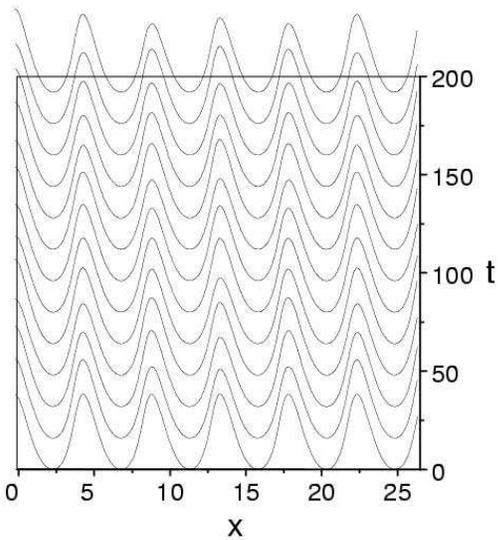}

\caption{ Time evolution of the first component of the cn-cn
solution reported in Fig.\ref{FIG1}. To check stability the
solution was slightly modulated in space with a profile of the
form $0.001 \cos(.01 x)$. A similar plot is obtained for the
second component in Fig. \ref{FIG1}. Parameters are fixed as in
Fig1. \label{FIG2}}
\end{figure}

The possibility that localized states of soliton type can be
generated from modulational instability of Bloch states at the
edge of the Brillouin zone, was analytically and numerically
proved for a single component BEC in optical lattice,  in the
cases of one \cite{konsa}, two and three spatial dimensions
\cite{bks}. In order to explore the same possibility to occur also
in the present periodic two-components system we recourse to
numerical simulations. To this regard we have used an operator
splitting method using fast Fourier transform to integrate Eqs.
(\ref{eq:1}), (\ref{eq:1'}) with initial condition taken as one of
the exact solutions derived above modulated by a long wavelength
$L$ ($2 \pi/k\ll \pi/L$) and small amplitude sinusolidal profile.

In Fig.1 we depict the initial profiles of the two components
cn-cn solution plotted against the potential profile, while in
Fig.2 we show the time evolution of the first component of this
solution in presence of a small modulation of the type $.001
\cos(.01 x)$.

\begin{figure}[h]
\includegraphics[scale=0.45,clip]{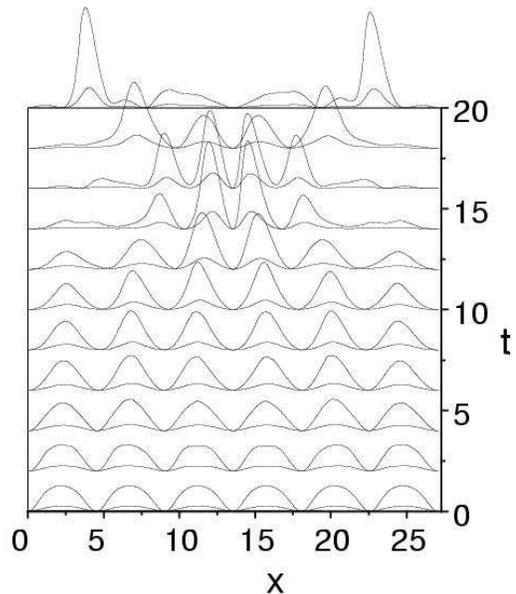}

\caption{Time evolution of the unstable sn-sn solution (notice
that both components of the solution are depicted at each time).
The initial amplitudes are taken $\gamma_1=- 0.41404$, and
$\gamma_2= 0.92582$. Parameters are fixed as in Fig. \ref{FIG1}
and the modulational initial profile is taken as in
Fig.\ref{FIG2}. Notice the emergence of coupled soliton
components out of the instability. \label{FIG3}}
\end{figure}

\begin{figure}[h]
\includegraphics[scale=0.45,clip]{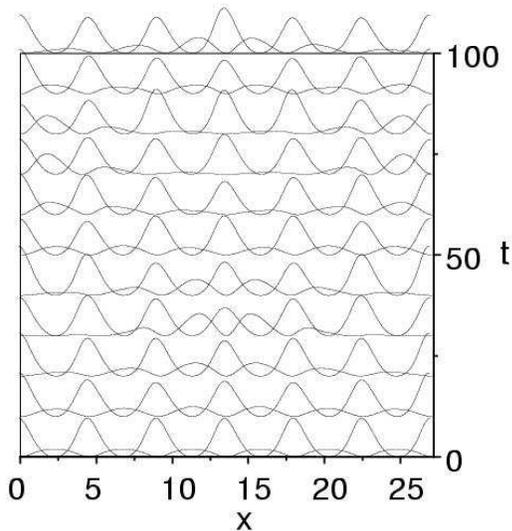}

\caption{ Same as in Fig.\ref{FIG3} but for the unstable sn-cn
solution. The initial amplitudes are taken as $\gamma_1=- 0.41404
$, and $\gamma_2= 0.92582$. Parameters are fixed as in
Fig.\ref{FIG1}. Notice that the cn component is larger and more
stable. At t=100 a bright-dark soliton is formed in the center.
\label{FIG4} }
\end{figure}

We see that the profile remains stable for long time (the same is
true also for the other component) indicating that the cn-cn
solution is stable against small modulations. We find that, except
for this, all the other solutions display modulational
instabilities out of which localized states emerge. This is
clearly seen in Fig.3 where the time evolution of the unstable
sn-sn solution is reported. Notice that in contrast with Fig.2
instability develops very quickly (already at time t=10), out of
which two components bright soliton states emerge,  as clearly
seen  at time $t=20$ (notice that the bright soliton consists of
two coupled solitons (one for each component) one bigger than the
other.

\begin{figure}[h]
\includegraphics[scale=0.45,clip]{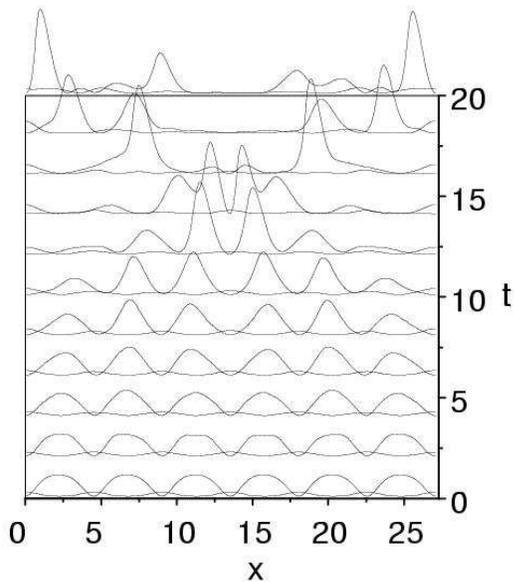}

\caption{Same as in Fig.\ref{FIG3} but for the unstable sn-cn
solution. The initial amplitudes are taken as $\gamma_1=
0.92582$, and $\gamma_2= - 0.41404$. Parameters are fixed  as in
Fig.\ref{FIG1}. Notice that the unstable sn component dominates
and soliton generation is more effective. \label{FIG5}}
\end{figure}

In Fig.4 and Fig. 5 we show the time evolution of unstable sn-cn
solutions with different amplitude ratio of the sn and cn
components. In Fig.4 the stable cn component is larger than the
unstable sn one, while in Fig. 5 we have the opposite. We see
that, although in both cases we have instability, the case with
larger stable component is obviously much more stable and  less
effective in creating localized states than the other. Also notice
from Fig. 4 that a small amplitude localized state is formed from
the sn component at time t=100 in the middle of the line which
seems to have a character opposite to the one of the other
component. We see indeed  that when matter density is higher in
one component it is lower in the other, this suggesting  a sort of
bright-dark coupling.

A better characterization of all possible states  arising from the
mixing of stable and unstable components requires a more accurate
analysis. It is interesting to investigate also solutions
involving dn components since these, in contrast with sn and cn
components, have non zero spatial average, i.e. they are periodic
waves on top of a constant background.
\begin{figure}[h]
\includegraphics[scale=0.45,clip]{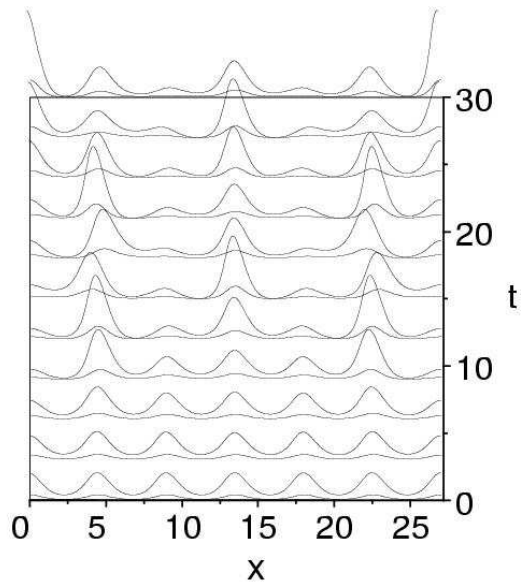}

\caption{Same as in Fig.\ref{FIG3} but for the unstable dn-dn
solution. Initial amplitudes are $\gamma_1= -0.46291$, and
$\gamma_2=1.035099$. Other parameters are fixed as in
Fig.\ref{FIG1}.\label{FIG6} }
\end{figure}

In Fig.6 we depict the time evolution of a dn-dn solution from
which we see that it is modulationally unstable, leading to the
formation of brigth solitons of the same type observed for the
sn-sn case. In Figs.7-8 we depict similar evolutions for the cases
dn-sn and dn-cn. Also in this case we observe that the mixing with
the unstable sn component is more effective than the one with the
stable cn component in creating localized excitations of soliton
type (the three brigth solitons formed in Fig.7 at time $t\approx
10$ remain equally spaced and well localized also on longer
times). By increasing the cn component of the dn-cn solution of
Fig.8, we also find that the time evolution becomes more stable,
as discussed for the sn-cn case.

This analysis shows that the exact trivial phase solutions of the
previous section are very useful to create localized excitations
of two components BEC in optical lattice. The fact that these
solutions are Bloch states at the edge of the  Brillouin also
suggests a way to create them in a real experiment. One could
indeed start from a uniform density distribution of the matter in
the potential wells, and accelerate the lattice until the state
reaches the edge of the Brillouin zone where the modulational
instability take place.

\begin{figure}[ht]
\includegraphics[scale=0.45,clip]{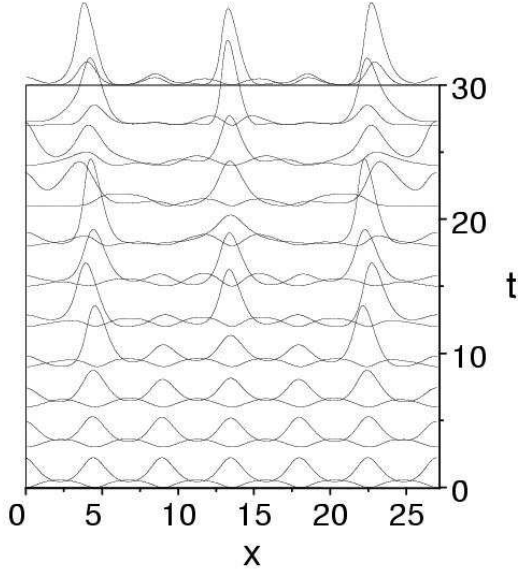}

\caption{Same as in Fig.\ref{FIG3} but for the the dn-sn
solution. Initial amplitudes are  $\gamma_1= -0.92582$, and
$\gamma_2=0.46291$. Other parameters are fixed as in
Fig.\ref{FIG1}.\label{FIG7} }
\end{figure}

\begin{figure}[ht]
\includegraphics[scale=0.45,clip]{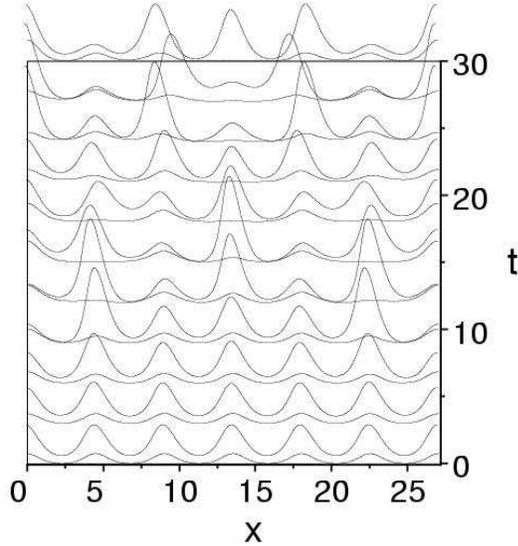}

\caption{Same as in Fig.\ref{FIG3} but for the dn-cn solution.
Initial amplitudes are  $\gamma_1= 0.46291$, and
$\gamma_2=-0.92582$. Other parameters are fixed as in
Fig.\ref{FIG1}.\label{FIG8} }
\end{figure}

\section{Discussions}\label{sec:Dis}

In this section we briefly discuss the above results in
comparison with those of Ref \cite{Deco} in which an $n$-component
NLS-type equation with external potential, whose
strength can be different for each component, was also considered.
Changing somewhat notations to avoid confusion with ours we write
it down in the form:
\begin{eqnarray}\label{eq:NLS-n}
i {\partial \psi _j  \over \partial t } &=& - {1  \over 2\mu _j }
{\partial^2 \psi _j  \over \partial x^2 } + V_j(x) \psi _j +
\sum_{p=1}^{n} a_{jp} |\psi _p|^2 \psi _j,  \\
V_j(x) &=& -V_{0j} \sn^2(\alpha x,k), \qquad j=1,\dots,n.
\end{eqnarray}

In Ref. \cite{Deco} three types of trivial phase solutions
are analyzed in more details for a rather special ansatz for $\Psi _j $;
in fact it is required there that all
$\Psi _j $ up to a standard phase factor are proportional to the same
function $\psi (x,k)$.  This means that the systems of $n $ equations will
reduce to just one equation for $\psi (x,k) $; the remaining $n-1 $
equations must follow as a consequence of the first one and a set of
constraints on the coefficients $a_{jl} $, $N_j $, $\mu _j $,
$\omega _j $ in the notations of \cite{Deco}. The same argument holds true
also for three of our solutions, cases 6, 7 and 8 which we added just for
the sake of completeness.

The Hamiltonian of the $n $-component NLS (\ref{eq:NLS-n}) is:
\begin{eqnarray}\label{eq:H}
H&=& \int_{}^{} dx \left[ \sum_{k=1}^{n} {1 \over 2\mu _j } \left| {
\partial \psi _j \over \partial x }\right|^2 + {1  \over 2 }
\sum_{j,p=1}^{n} a_{jp} |\psi _j|^2 |\psi _p|^2 \right. \nonumber\\
&+& \left. \sum_{j=1}^{n} V_j(x) |\psi _j|^2 \right].
\end{eqnarray}
where the integration goes over one period.
Let us now assume that our solution is of the form:
\begin{equation}\label{eq:1-c}
\psi _j(x) = n_j(x,t) \psi (x,t), \qquad n_j(x,t) = e^{-i\omega
_jt + i \Theta _j(x)} \sqrt{N_j}
\end{equation}
where $N_j>0 $ and $\Theta _j(x) $ appears only in the non-trivial phase
case and is determined by:
\begin{equation}\label{eq:The}
{d\Theta _j  \over dx } = {C_j  \over N_j |\psi (x)|^2 }.
\end{equation}
Inserting (\ref{eq:1-c}) into the Hamiltonian we easily get the following
reduced Hamiltonian:
\begin{eqnarray}\label{eq:H_r}
H_{\rm red} &=& \int_{}^{}dx\, \left[ {1 \over 2} M_0 \left|
{\partial \psi \over \partial x} \right|^2 + {1  \over 2 } M_{-1}
{1  \over |\psi |^2 } + V(x) |\psi |^2 \right. \nonumber\\
 &+& \left. {1  \over 2 } W_0 |\psi |^4 \right], \\
M_0 &=& \sum_{j=1}^{n} {1  \over N_j\mu _j }, \qquad M_{-1} =
\sum_{j=1}^{n} {N_j  \over \mu _j } , \nonumber\\
V(x) &=& v_0 \sn^2(\alpha x,k), \qquad W_0 = \sum_{j,p=1}^{n}
a_{jp} N_jN_p.\nonumber
\end{eqnarray}
which describes the dynamics of the effective field $\psi (x,t) $.
The result for the trivial phase solution case leads to $H_{\rm red} $
with $M_{-1}=0 $. That is why we believe that the multicomponent effects
should be analyzed by using ansatz more general than (\ref{eq:1-c}).

At the same time an important result of \cite{Deco} is the detailed,
both analytical and numerical, analysis of type B solutions to
(\ref{eq:NLS-n}) though again using (\ref{eq:1-c}).

In the present paper we considered intrinsic two-component
solutions i.e. solutions  with different amplitudes. These
solutions seems to have stability property which are not trivial
consequence of the theorems proved in \cite{Deco} and deserve
additional studies. Besides enlarging the set of solutions,  we
have also shown the role played by these solutions as initial
states from which localized matter waves (solitons) can be
generated through the  modulational instability mechanism.

\section{Conclusions}
In conclusion, we have considered the two-component CNLS
with an elliptic potential as a model for trapped,
quasi-one-dimensional two-component BECs. Classes of elliptic,
solitary wave and trigonometric solutions have been presented.

From a physical point of view the solutions discussed in this
paper are exact nonlinear Bloch states with wave number at the
edges of the Brillouin zone. These solutions, except for the
$cn-cn$ one, are unstable under small amplitude and large
wavelength modulations. Two component matter solitons arise from
these unstable solution via a modulational instability mechanism
which resemble the one described in Refs. \cite{konsa,bks} for
single component BEC in optical lattices.

Further perspectives of finding stable periodic solutions to the
2-component problem could be linked to investigations of
finite-gap solutions of Manakov system given in terms of
multi-dimensional $\theta$-functions \cite{ahh93} and \cite{eei03}
and to reduction of finite-gap solutions to elliptic functions
\cite{book}. Interesting classes of periodic solutions can be also
obtained as the result of reduction of the Manakov system to
completely integrable two-particle system interacting with fourth
order potential \cite{ceek95,ceek00} and \cite{EilEnoKo}.

\acknowledgements { V.V.K. acknowledges Prof. Deconinck for
discussing results in Ref.\cite{Deco} prior their publication.}
V.S.G. and V.Z.E. wish to thank the Department of Physics
"E.R.Caianiello" for the hospitality received, and  the
University of Salerno for providing an eight months research
grant during which most of this work was done. V.V.K acknowledges
support from the European grant, COSYC n.o HPRN-CT-2000-00158.
M.S. acknowledge partial financial support from the MIUR, trough
the inter-university project PRIN-2000, and from the European
grant LOCNET grant no. HPRN-CT-1999-00163.


\end{document}